\documentclass[pra,%
 reprint,
 amsmath,amssymb,
 aps,
]{revtex4-2}

\usepackage{amsmath,amssymb,amsfonts}
\usepackage{algorithmic}
\usepackage{graphicx}
\usepackage[caption=false]{subfig} 
\usepackage{tabularx}
\usepackage{textcomp}
\usepackage{xcolor}
\usepackage{color}
\usepackage{algorithm,algorithmic} 
\usepackage{tikz}
\usepackage{nicematrix}
\NiceMatrixOptions{cell-space-top-limit = 1pt,cell-space-bottom-limit = 1pt}
\usepackage{dcolumn}
\usepackage{bm}
\usepackage{easyReview} 

\def\BibTeX{{\rm B\kern-.05em{\sc i\kern-.025em b}\kern-.08em
    T\kern-.1667em\lower.7ex\hbox{E}\kern-.125emX}}

\begin{document}

\preprint{APS/123-QED}

\title{Robust Quantum State Generation in Symmetric Spin Networks}

\author{André Luiz P. de Lima}
\author{Jr-Shin Li}%
\affiliation{%
 Department of Electrical and Systems Engineering\\ Washington University in St. Louis
}%

\author{Luke S. Baker}
\author{Anatoly Zlotnik}
\affiliation{
 Applied Mathematics and Plasma Physics Group\\ Los Alamos National Laboratory
}%

\author{Andrew K. Harter}
\author{Michael J. Martin}
\affiliation{%
 Quantum Technologies Team\\ Los Alamos National Laboratory
}%




\date{\today}

\begin{abstract}
   In this work, we consider a parameterized Ising model with long-range symmetric pairwise interactions on a network of spin $\frac{1}{2}$ particles. The system is designed with symmetric dynamics, allowing for the reduction of the state space to a subspace defined by the set of Dicke states. We propose a method for designing robust electromagnetic amplitude pulses based on a moment quantization approach. The introduced parameter accounts for uncertainties in the electromagnetic field, resulting in a family of distinct Hamiltonians. By employing a discretized moment-based quantization technique, we design a control pulse capable of simultaneously steering an infinite collection of dynamical systems to compensate for parameter variations. This approach benefits from the duality between the infinite-dimensional parameterized system and its finite-dimensional trucnated moment dynamics. Simulation results demonstrate the efficacy of this method in achieving states of significant interest in quantum sensing, including the GHZ and W states.
\end{abstract}

\maketitle


\section{INTRODUCTION}

Significant advancements in detection and estimation methodologies have emerged through the integration of quantum mechanics principles and technologies \cite{Hel:69, Guo:22}. Quantum metrology, in particular, has revolutionized the development of innovative techniques for measuring highly elusive physical phenomena, such as gravitational waves, dark matter, and other fundamental properties \cite{Gio:11, Tot:14}. One of the most distinct advantages of employing quantum systems in these settings is the ability to achieve precision beyond that of any classical approach. 

In classical statistics, the concepts of Fisher information and the Cramer-Rao bound establish the minimum variance that an estimator can achieve, serving as an optimal benchmark for parameter estimation problems \cite{Seng:95, Poor:13}. While these concepts remain relevant in quantum estimation, they can be exceeded due to the phenomenon of entanglement among quantum agents. This leads to the introduction of Quantum Fisher Information (QFI), which adheres to an enhanced optimal bound dictated by the Heisenberg Limit (HL) \cite{Hua:24, Xio:10}. The resultant improvement in estimation precision positions quantum metrology as a highly advantageous and compelling field for achieving unprecedented accuracy in physical measurements.

However, the generation of entangled states, which is crucial for attaining the HL, is a complex endeavor. The challenge is compounded by the fact that the type of entangled state is critical for reaching the HL, and the problem's complexity escalates exponentially with the addition of new quantum elements \cite{Rub:00, Erh:20}. A viable strategy to manage these challenges is to define a system that spans a subset of states with different scaling behavior as agents increase, while incorporating entangled states capable of reaching the HL. For example, in phase estimation, a quantum system can be characterized by its set of Dicke states, which scales linearly in dimension as new agents are added and can achieve the HL \cite{Fri:07, Sal:24, Gro:12}. This has led to the study of highly symmetric networks represented by Dicke states, with proven operator controllability for a system evolving according to the Ising model with long-range pairwise interactions \cite{Alb:18, Chen:20}. This has also led to the proof that the subspace of symmetric quantum states can also lead to robustness toward erasure and detuning errors in quantum metrology applications \cite{Ouy:21}.

Despite the effective dimensionality reduction and robustness with respect to decoherence that can be 
achieved through the adoption of symmetric subspaces, designing pulses robust to empirical irregularities remains a highly challenging problem. Methods to enhance fidelity under such conditions often rely on adiabatic approaches for gradual state preparation \cite{Car:24}, sequential pulse emission with carefully designed spin rotations \cite{Sto:23}, or asymptotically convergent dynamic design \cite{Ran:18}. While these solutions have demonstrated their effectiveness, the application of optimization concepts in control pulse design remains largely unexplored. Incorporating such techniques could lead to significant improvements in overall fidelity.

In quantum control applications, this challenge is frequently addressed by considering a parameterized dynamical system and simultaneously optimizing control designs. To this end, sampling-based approaches employing large-scale optimization algorithms are commonly used for a subset of the ensemble, with the GRAPE (Gradient Ascent Pulse Engineering) method being a prominent example \cite{Say:20, Say:23, Pet:23}. However, these sampling approaches inherently involve a trade-off between computational cost and efficiency due to the scaling complexities inherent in the task.

Fortunately, the parameterized nature of similar problems has given rise to the field of ensemble control, which has introduced methodologies to enhance the efficacy of existing optimization techniques. Within this field, transformation tools, such as polynomial-based moment states for dual system representations, have been developed \cite{Zeng:16, Nar:20} and successfully applied to both quantum-related applications \cite{De:24, Ning:22, Li:22} and large-scale systems in general \cite{Vu:24, De:24_b}.

In this study, we focus on the design of robust pulses for an Ising system under parameterized dynamics to generate symmetric quantum states. Specifically, we target the preparation of states of significant interest in quantum estimation, such as GHZ and W states. The analyzed system, previously employed in metrological applications \cite{Car:24, Hua:24, Per:20}, features a deterministic entangling Hamiltonian that implements an effective one-axis twisting (OAT) interaction, alongside two control Hamiltonians acting on orthogonal axes. These control Hamiltonians are parameterized to capture the heterogeneous dynamics of the system.  We then formulate an optimal control problem that is used to steer such ensembles into several coherent states of interest, and explore its performance.

The organization of this paper is as follows: Section \ref{sec:Network_model} introduces the quantum system under study, including the symmetric subspace of states and the Ising model Hamiltonian governing its evolution. Section \ref{sec:Ensemble_Moment} presents the ensemble interpretation of the problem and details the Legendre moment representation of the system, which becomes central to the control approach. In Section \ref{sec:Algorithm}, we describe the application of the moment model in the development of an optimal control law and algorithm for pulse design. Finally, Section \ref{sec:Results} provides our results on robust control design for Hamiltonian controls in the presence of inhomogeneities, followed by concluding remarks in Section \ref{sec:Conclusion}.

\section{SYMMETRIC SPIN NETWORK OF QUANTUM PARTICLES}\label{sec:Network_model}

We consider a network comprised of $N$ total $\frac{1}{2}$-spin quantum particles, steered and entangled by a collection of electromagnetic fields. A primary challenge in representing this system lies in the dimensionality issue, stemming from the exponential growth of the Hilbert space with the number of particles, with dimension equal to $2^{N}$ \cite{Nov:15}. To ensure tractability in studying these networks and their applications, it is common to define an invariant subspace that achieves significant dimensional reduction while remaining practical for realistic reproduction \cite{Car:24}. One such approach involves describing a symmetric set of states evolving under dynamics dictated by correspondingly symmetric Hamiltonians, which will be the focus of this work.

\subsection{Symmetric Dicke States and Coveted Quantum Profiles}\label{subsec:Dicke_States}

The proposed reduced subspace evolves under a basis of highly symmetric quantum states. One such basis is defined by the Dicke states, which has been used previously in applications regarding quantum metrology, where it was shown that it was able to define states that surpasses the Standard Quantum Limit (SQL), being able to achieve the HL \cite{Sal:24, Xio:10}. The Dicke states are able to exactly describe the number of excited particles in a population of two-level atoms following Bloch dynamics \cite{Fri:07}. These are represented using the Dicke basis for symmetric spin particles states, which is defined as


\begin{equation}\label{eq:Dicke_States}
    |S, m\rangle = \binom{N}{S+m}^{-\frac{1}{2}} \sum_{\phi_{i} \in \Phi_{S+m}^{N}} |\phi_{i} \rangle
\end{equation}
where $S=\frac{N}{2}$, $m \in \mathcal{M} := \{m=S, S-1, \dots, -S\}$, and $\Phi_{S+m}^{N}$ denotes the set of all permutations of $S+m$ excited spins ($|\uparrow\rangle$) and $S-m$ unexcited spins ($|\downarrow \rangle$) among the set of $N$ particles that compose the network. Besides being constrained to the set of symmetric states, there are various generable profiles that are of significance in quantum metrology. Specifically, the coherent states that we examine here are
\begin{equation*}
    \begin{split}
        |W\rangle & =  |S, -(S-1)\rangle,\\
        |HEDS\rangle & =  \begin{cases}
			|S, 0\rangle, & \text{if $N$ even},\\
            |S, -\frac{1}{2}\rangle, & \text{otherwise},
		 \end{cases}\\
        |GHZ\rangle & =  \frac{1}{\sqrt{2}}(|S, S\rangle + |S, -S\rangle).
    \end{split}
\end{equation*}
Besides representing fairly different configurations, all of these states are relevant in quantum metrology applications. The $|W\rangle$ state is used in communication applications due to its inherent robustness towards loss of information \cite{Dur:00}. The Greenberger-Horne-Zeilinger $|GHZ\rangle$ state is used in error correction due to promoting the maximum entanglement among particles \cite{Hua:24} and the Highly Excited Dicke State $|HEDS\rangle$ promotes the greatest increment in QFI, reaching the HL \cite{Car:24}.

\subsection{Dynamical Evolution of Network Following Ising Model}\label{subsec:Ising_Model}

Frequently regarded as one of the standard models for describing quantum field phenomena, the Ising model offers a straightforward framework to represent specific processes involving a lattice of qubits with entangling dynamics. These dynamics are foundational in various applications, including quantum computing, metrology, and materials science \cite{Car:24, Gus:19, Wolf:00}. The Hamiltonian governing this system is expressed as follows:

\begin{equation} \label{eq:Hamiltonian}
    H = \chi H_{zz} + H_{x}u_{x}(t) + H_{z}u_{z}(t),
\end{equation}

where

\begin{equation*}
    \begin{split}
        H_{ii}= & \sum_{1\leq j < k \leq N} \sigma_{i}^{j}\sigma_{i}^{k},
    \end{split}
\end{equation*}
\begin{equation*}
    \begin{split}
        H_{i}= & \sum_{1\leq j \leq N} \sigma_{i}^{j},
    \end{split}
\end{equation*}
where $\sigma_{i}^{j}$ denotes the Pauli operators for $i = x, y, z$ acting on the $j^{th}$ spin in the network. The dynamics for this system can be divided into two components: the drift term $H_{zz}$, which governs particle entanglement, and the control Hamiltonians $H_{x}$ and $H_{z}$, which enable universal rotations of particles along the $x$- and $z$-axes, respectively. The shearing parameter $\chi$ is used as a scaling factor in the simulations, and the functions $u_{x}(t)$ and $u_{z}(t)$ are time-varying functions that we suppose can be used as controls. 


By adopting the Dicke basis described in Section \ref{subsec:Dicke_States}, we can represent the evolution of the system states under the Hamiltonian in equation \eqref{eq:Hamiltonian}. Due to the symmetric nature of the Hamiltonian, the Dicke basis defines an invariant subspace, as the unitary operators derived from the dynamics of the system are permutation-invariant \cite{Alb:18}.  Consequentially, the equation of motion for the probability amplitude can be derived from the aforementioned Hamiltonian. This leads to a closed differential equation governing the evolution of the probability amplitudes, i.e.,

\begin{multline}\label{eq:Prob_amp}
    i\dot{C}_{m}(t) = \left[ \chi (2m^{2}-N/2) + 2mu_{z}(t) \right] C_{m}(t) \\ + {u_{x}(t)}\left( \zeta_{+}C_{m+1}(t) + \zeta_{-}C_{m-1}(t) \right),
\end{multline}
where $C_{m}(t)$ represents the probability amplitude for the corresponding eigenstate $|S, m\rangle$ at the time $t \in [0, T]$, and $\zeta_{\pm} = \sqrt{(S \mp m)(S \pm m + 1)}$. Within this framework, the eigenstates corresponding to the target profiles defined in Section \ref{subsec:Dicke_States} can also be characterized by their probability amplitudes, 
\begin{equation*}
    \begin{split}
        \lVert C_{-(S-1)}^{W} \rVert_{2}^{2} & =  1,\\
        \lVert C_{m_{0}}^{HEDS}\rVert_{2}^{2} & =  1,\quad m_{0}=0\text{ or }m_{0}=-\frac{1}{2},\\
        \lVert C_{m_{S}}^{GHZ}\rVert_{2}^{2} & =  \frac{1}{{2}},\quad m_{S}=S\text{ or }m_{S}=-S.
    \end{split}
\end{equation*}
These profiles will be reference points for the design of robust control pulses, which will be detailed in Section \ref{sec:Algorithm}.

\section{Ensemble System Interpretation and its Moment Kernel}\label{sec:Ensemble_Moment}

Although the Ising model described in Section \ref{subsec:Ising_Model} provides a dynamical framework for the symmetric spin network, it is inherently affected by uncertainties associated with controlling the system. Fortunately, such uncertainties are often modeled as parameters defined within confidence intervals, characterizing the evolution of a continuous function over time. This approach has enabled previous studies to decompose the vector space of analogous ensembles using a polynomial basis, defining an isomorphism with an equivalent ensemble of moments \cite{Nar:24}. By leveraging the duality between these representations and the unified framework provided by the moments, we propose the robust design of control pulses through direct engagement with the corresponding moment dynamics.

\subsection{Parameterized Ensemble Ising Model}

Due to the intrinsic uncertainty in the electromagnetic control pulse, the Hamiltonian 
in equation \eqref{eq:Hamiltonian} can be reformulated by introducing time-invariant parameters $\xi \in [\xi_{min}, \xi_{max}]$ and $\zeta \in [\zeta_{min}, \zeta_{max}]$, leading to the following expression:

\begin{equation} \label{eq:Hamiltonian_Ensemble}
    H(\xi, \zeta) = \chi H_{zz} + \xi H_{x}u_{x}(t) + \zeta H_{z}u_{z}(t)
\end{equation}

The Hamiltonion $H(\xi, \zeta)$ defined in equation \eqref{eq:Hamiltonian_Ensemble} represents a continuum of dynamics encapsulating the potential evolutions of the studied Ising system. Given the nature of control inhomogeneity, this uncertainty is often attributed to noise or variability in electromagnetic intensity, which is typically expressed as a percentage. This allows the compact intervals to be redefined as $[\xi_{min}, \xi_{max}] = [1-\delta_{\xi}, 1+\delta_{\xi}]$ and $[\zeta_{min}, \zeta_{max}] = [1-\delta_{\zeta}, 1+\delta_{\zeta}]$, in which $\delta_{\xi}, \delta_{\zeta} \in [0, 1)$.

We can further expand the dynamical representation of this system by expressing equation \eqref{eq:Prob_amp} within the framework of the ensemble system perspective, as 
\begin{multline}\label{eq:Prob_amp_Ensemble}
    i\dot{C_{m}(t, \xi, \zeta)} = \left[ \chi (2m^{2}-N/2) + 2m\zeta u_{z}(t) \right] C_{m}(t) \\ + \xi{u_{x}(t)}\left( \zeta_{+}C_{m+1}(t) + \zeta_{-}C_{m-1}(t) \right)
\end{multline}
For the remainder of this paper, we use equation \eqref{eq:Prob_amp_Ensemble} and its corresponding moment ensemble dual representation. Moreover, without loss of generality, we will assume $\chi = 1$, which is dynamically interpreted as the units of time scaled by a factor $\chi$.

\subsection{Moment Kernel with a Legendre Polynomial Basis}

The challenging aspects of addressing the parameterized equation \eqref{eq:Prob_amp_Ensemble} stem from the need to describe a continuum of vector spaces, resulting in an uncountable infinite-dimensional system. To mitigate this complexity, the moment kernel has been proposed as an effective tool. Leveraging the insights provided by the Stone-Weierstrass theorem \cite{Fol_99}, this method introduces a discrete set of moments that, collectively, provide a complete representation of the system ensemble. Assuming a function $f(\cdot)$ of an arbitrary parameter $\varepsilon$, defined on a separable Hilbert space, it is possible to define a polynomial basis $\{\rho(\varepsilon) \}_{i=0}^{\infty}$ such that this function can be decomposed into its respective moments, expressed through the 
inner product
\begin{equation}\label{eq:inner_product}
    \mathbf{m}_{k} = \langle f(\varepsilon), \rho(\varepsilon) \rangle.
\end{equation}
Previous studies have identified certain desirable properties for selecting a polynomial basis, with orthogonality and completeness over a compact interval being among the most significant \cite{Li:22}. A commonly preferred choice is the normalized set of Legendre polynomials $\{P(\varepsilon) \}_{i=0}^{\infty}$, which form an orthogonal system over the interval $[-1, 1]$, satisfying $\langle P{n}(\varepsilon), P_{m}(\varepsilon) \rangle = \int_{-1}^{1} P_{n}(\varepsilon)P_{m}(\varepsilon)d\varepsilon = \delta_{nm}$. Utilizing this property, the moment states for the ensemble described by equation \eqref{eq:Prob_amp_Ensemble} can be defined as 
\begin{equation}\label{eq:Moment_Ensemble}
  \!\!\!\!  \mathbf{m}_{m, i, j}(t) \!=\! \int_{-1}^{1}\int_{-1}^{1} C_{m}^{*}(t, \xi^{*}, \zeta^{*})P_{i}(\xi^{*})P_{j}(\zeta^{*})d\xi^{*} d\zeta^{*},
\end{equation}
where $m_{m,i,j}$ is the moment related to the function $C_{m}(t, \xi, \zeta)$ of the $i^{th}$ and $j^{th}$ orders in relation to the parameters $\xi$ and $\zeta$, respectively. The function $C_{m}^{*}(t, \xi^{*}, \zeta^{*})$ is defined by a isomorphism with function $C_{m}(t, \xi, \zeta)$ such that $\xi^{*} = \frac{\xi-1}{\delta_{\xi}}$ and $\zeta^{*} = \frac{\zeta-1}{\delta_{\zeta}}$.

By differentiating equation \eqref{eq:Moment_Ensemble} with respect to time and substituting the differential equation governing $C_{m}(t, \xi, \zeta)$ from equation \eqref{eq:Prob_amp_Ensemble}, we can derive the dynamical evolution of the ensemble in moment space. The resulting moment evolution law is 

\begin{equation}\label{eq:Moment_Ensemble_Dynamics}
\!\!\!\!    \dot{\mathbf{m}}_{m, i, j}(t) \!=\! \int_{-1}^{1}\int_{-1}^{1} \dot{C}_{m}^{*}(t, \xi^{*}, \zeta^{*})P_{i}(\xi^{*})P_{j}(\zeta^{*})d\xi^{*} d\zeta^{*}.
\end{equation}
Finally, we emphasize that a proven isometric isomorphism exists between the moment space and the ensemble vector space. This fundamental relationship allows us to formulate optimization problems using the moment states, ensuring that the resulting solutions achieve equivalent objectives in the original ensemble representation.

\section{Optimal Pulse Design Through Iterative Quadratic Algorithm}\label{sec:Algorithm}

We design robust pulses for this system using an iterative quadratic programming approach previously employed in similar robust control problems \cite{Bak:24}. This method is specifically tailored for bilinear systems, where the dynamics are linearized and evolved recursively over small time intervals. While this approach can be computationally intensive due to the large scale of the problem, the polynomial decomposition introduced in Section \ref{sec:Ensemble_Moment} significantly reduces the dimensionality, making the problem more tractable.

To implement the control design, we formulate a constrained optimization problem as defined in equation \eqref{eq:moment_optimization}. In this control law, constraints arise from the system dynamics (as established in equation \eqref{eq:Moment_Ensemble_Dynamics}) and the physical limitations of the experimental setup. We refer to the latter limitations as the signal restrictions
\begin{subequations} \label{eq:sigrest}
\begin{align}
\!\!\!\!  u_x^{\min} & \leq u_x(t) \leq u_x^{\max}, \, u_z^{\min} \leq u_z(t) \leq u_z^{\max}, \\
\!\!\!\!  \Delta u_x^{\min} & \leq \dot{u}_x(t) \leq \Delta u_x^{\max}, \, \Delta u_z^{\min} \leq \dot{u}_z(t) \leq \Delta u_z^{\max}.
\end{align}
\end{subequations}

Suppose that we aim to achieve a final state $|\psi_{f}\rangle$, which corresponds to one of the previously defined target states ($|W\rangle$, $|HEDS\rangle$ or $|GHZ\rangle$). We then define the set $\mathcal{A}:= \{a \in \mathcal{M} \left|\hspace{0.5em} |\langle | \psi_{f}| S, a \rangle| >0 \}\right.$ and the element $a_{max} = \sup(\mathcal{A})$, leading to the following formulation:

\small{\begin{equation} \label{eq:moment_optimization}
\begin{aligned}
    \!\!\!\!\!\!\!\!\!\! \min_{\substack{u_{x}(t)\\ u_{z}(t)}} \,\, & \bigg\lVert \sum_{\substack{i,j\\a\notin\mathcal{A}}} |\mathbf{m}_{a,i,j}(T)| +\!\!\!\! \sum_{\substack{i, j, a\in\mathcal{A} \\ a \neq a_{max}}} |\mathbf{m}_{a,i,j}(T) - 4\delta_{i0}\delta_{j0}\langle | \psi_{f}| S, a \rangle|  \bigg\rVert_{2}\\
    s.t. \,\, & \textrm{Dynamics in equation } \eqref{eq:Moment_Ensemble_Dynamics},\\
    \quad & \textrm{Signal restrictions in equations }  \eqref{eq:sigrest}.
\end{aligned}\!\!\!\!\!\!\!
\end{equation}
}
\normalsize{}
In the above optimization problem, moment states corresponding to eigenvectors $|S, a\rangle$ for which $a\notin\mathcal{A}$ are nullified, as their associated moment values are zero, enforcing the equality $C_{a}(T, \cdot, \cdot)=0$. For eigenvectors where $a\in\mathcal{A}$, we aim to design a control pulse such that the probability amplitudes $C_{a}(T, \cdot, \cdot)$ match the desired absolute values $|\langle | \psi_{f}| S, a \rangle|$. To achieve this, the second term in the optimization problem minimizes moments of order higher than zero, while approximating the moment $\mathbf{m}_{a, 0, 0}$ (related to the polynomial of order 0) to the absolute desired amplitude for the respective eigenvector $|S, a\rangle$. Notably, the eigenvector $|S, a_{max}\rangle$ is excluded from this minimization process. Because the probability amplitudes must sum to one, ensuring the correct values for all other eigenvectors inherently guarantees the correct final amplitude for $|S, a_{max}\rangle$. This formulation offers a computational advantage, streamlining the design of robust control pulses.

Before presenting the simulation results, we emphasize that our approach has been formulated considering $C_{m} \in \mathbb{C}$, which introduces computational challenges due to the complex nature of these variables—particularly in constrained optimization settings. To address this, we decompose the dynamics in equation \eqref{eq:Prob_amp}, expressing the state as $C_{m} = C_{m}^{\mathbb{R}} + C_{m}^{\mathbb{C}}i$, where $C_{m}^{\mathbb{R}}, C_{m}^{\mathbb{C}}\in \mathbb{R}$ and $\lVert C_{m}^{\mathbb{R}} +  C_{m}^{\mathbb{C}}\rVert_{2}^{2}\leq1$. This allows the system to be represented by the real-valued vector $[C_{m}^{\mathbb{R}}, C_{m}^{\mathbb{C}}]^{T}$. All subsequent equations in this paper are modified accordingly to reflect this representation.

\begin{figure*}[t!]
\begin{center}
\includegraphics[width=\textwidth]{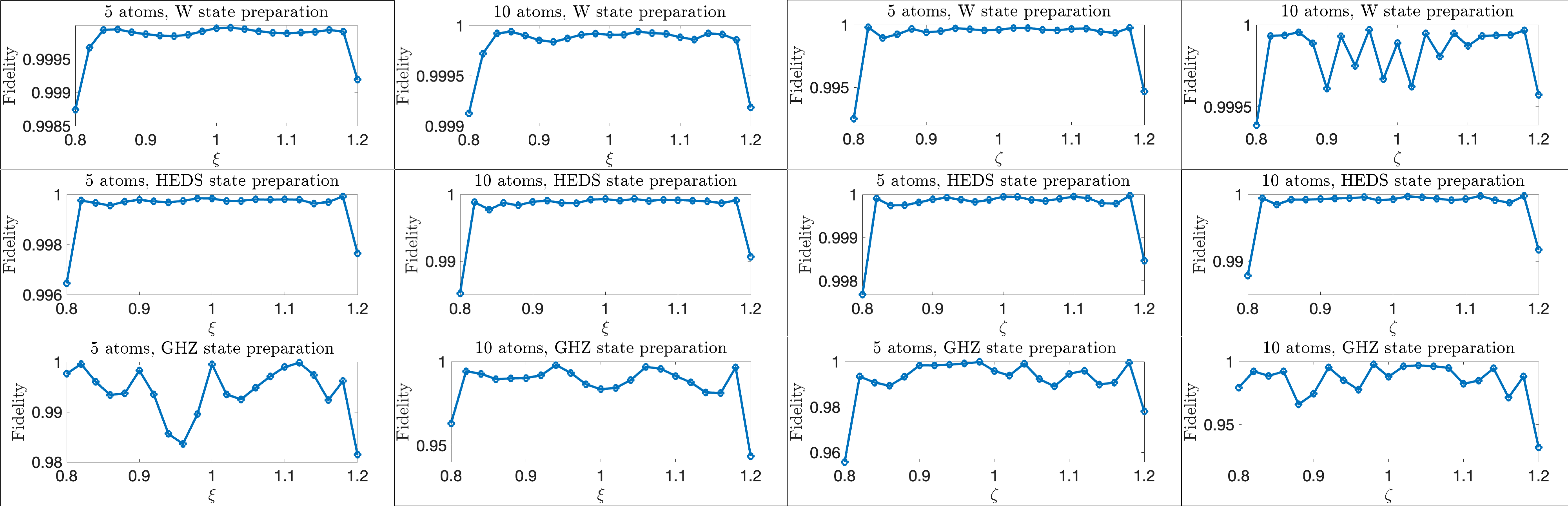}    
\vspace{-4ex}
\caption{Fidelity metric values for designed robust pulses. Designs are for $\xi\in[1-\delta_{\xi},1+\delta_{\xi}]$ with $\delta_{\xi} = 0.2$ for the first and second columns of plots, and $\zeta\in[1-\delta_{\zeta},1+\delta_{\zeta}]$ with $\delta_{\zeta} = 0.2$ for the latter columns. Desired states to achieve are, from top row to bottom, $|W\rangle$, $|HEDS\rangle$ and $|GHZ\rangle$. The experiments are also differentiated by the population of atoms, being equal to 5 for the first and third columns and equal to 10 for the second and fourth.}
\label{fig:Results_Fidelity}
\end{center}
\vspace{-4ex}
\end{figure*}

\begin{figure*}
\begin{center}
\includegraphics[width=\textwidth]{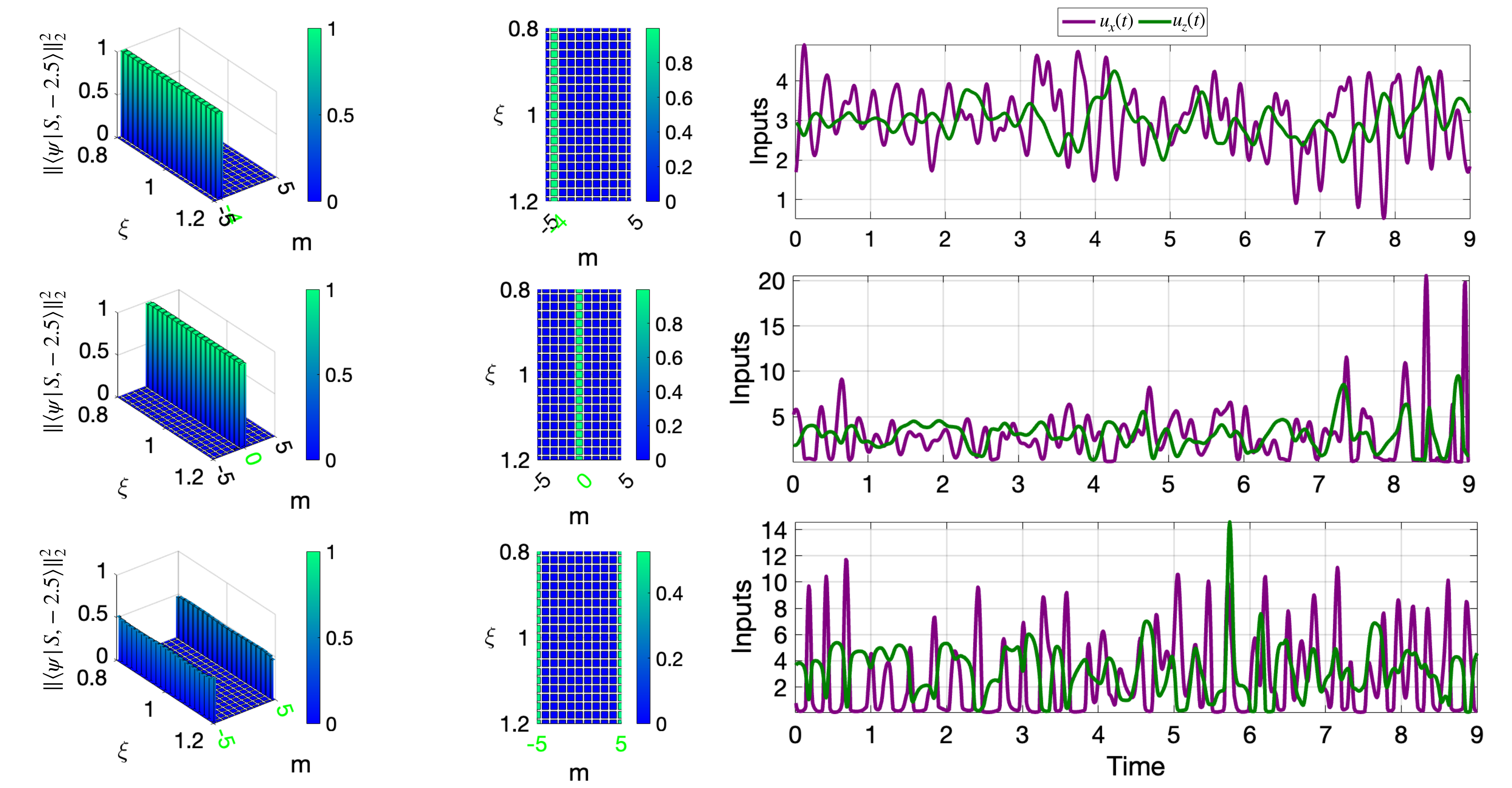}    
\vspace{-4ex}
\caption{Simulated results for the robust design of quantum states for a network with 10 atoms for a system with $\delta_{\xi} = 0.2$ and $\delta_{\zeta} = 0$. Results refer to the design of (from top to bottom) $|W\rangle$, $|HEDS\rangle$ and $|GHZ\rangle$ states. The left plots refer to the final achieved states as a function of the probability amplitude of $|S, m \rangle$ (represented here as $\lVert|S, m\rangle\rVert ^{2}_{2}$). The second plot is the same as the first plot viewed from above. The right plot shows the control profiles obtained for $u_{x}(t)$ and $u_{z}(t)$ for the total time $T=9$.}
\label{fig:Results_Simulations}
\end{center}
\vspace{-5ex}
\end{figure*}

\section{Simulation Results}\label{sec:Results}

The objective of the control design is to achieve robust pulse generation for the three key quantum metrology states, namely, $|W\rangle$, $|HEDS\rangle$ and $|GHZ\rangle$. The performance of these approaches is assessed using the fidelity metric $\mathcal{F}(\xi, \zeta, |\psi_{f}\rangle)$, as defined by
\begin{equation}\label{eq:Fidelity}
    \mathcal{F}(\xi, \zeta, |\psi_{f}\rangle) = 1-\sum_{a\in\mathcal{A}} \lVert |C_{a}(T, \xi, \zeta)| - \langle \psi_{f} |S,a \rangle \rVert_{2}^{2}.
\end{equation}
High-performance control is indicated when $\mathcal{F}(\xi, \zeta, |\psi_{f}\rangle)$ approaches unity for all systems characterized by $\xi$ and $\zeta$.

We assess the robustness of our methodology by examining two distinct scenarios. First, we evaluate the impact of robustness on a single inhomogeneous parameter, independently analyzing the control pulse performance for both $\xi$ and $\zeta$ independently. Subsequently, we extend this analysis to a scenario where the control pulse is designed to be robust against variations in both parameters simultaneously, demonstrating the overall effectiveness of our approach in handling nonuniform control dynamics.

\subsection{Robust Optimization of a Single Control Hamiltonian} \label{subsec:Results_Single}

We implement the control design and simulation for a spin network with long-distance Ising interactions. The control design is conducted on a system of moments up to the fourteenth order, following the dynamics outlined in equation \eqref{eq:Moment_Ensemble_Dynamics}. Simulations are then applied to the ensemble of probability amplitudes, governed by equation \eqref{eq:Prob_amp_Ensemble}, using a sampled ensemble to evaluate performance quantified using the fidelity metric in equation \eqref{eq:Fidelity}. We suppose that the network is initially in the ground state, given by $|\psi_{0}\rangle = |S, -S\rangle$.

The problem is analyzed in two distinctive settings. In the first, we define $\delta_{\xi} = 0.2$ while $\delta_{\zeta} = 0$, and in the second, we define $\delta_{\xi} = 0$ while $\delta_{\zeta} = 0.2$. For each setting, simulations are performed considering populations of 5 and 10 $\frac{1}{2}$-spin particles. The total duration of the simulated dynamics is defined by the parameter $T=9$, with time units scaled by the shearing parameter $\chi$. The system's time evolution is evaluated iteratively using a time step of $\Delta t = 0.01$. The signal restriction constraint bound values for all simulations are set to $u_x^{\min}=u_z^{\min}=0$, $u_x^{\max}=u_z^{\max} = 40$, $\Delta u_x^{\min}=\Delta u_z^{\min} = -10^{4}/t$, and $\Delta u_x^{\max}=\Delta u_z^{\max} = 10^{4}/t$.The initial control profiles $u_{x,0}(t)$ and $u_{z,0}(t)$ used to initialize the control synthesis method are defined as constant functions over time, i.e., $u_{x,0}(t) = u_{z,0}(t) = 3$.

\begin{figure*}[t!]
\begin{center}
\includegraphics[width=\textwidth]{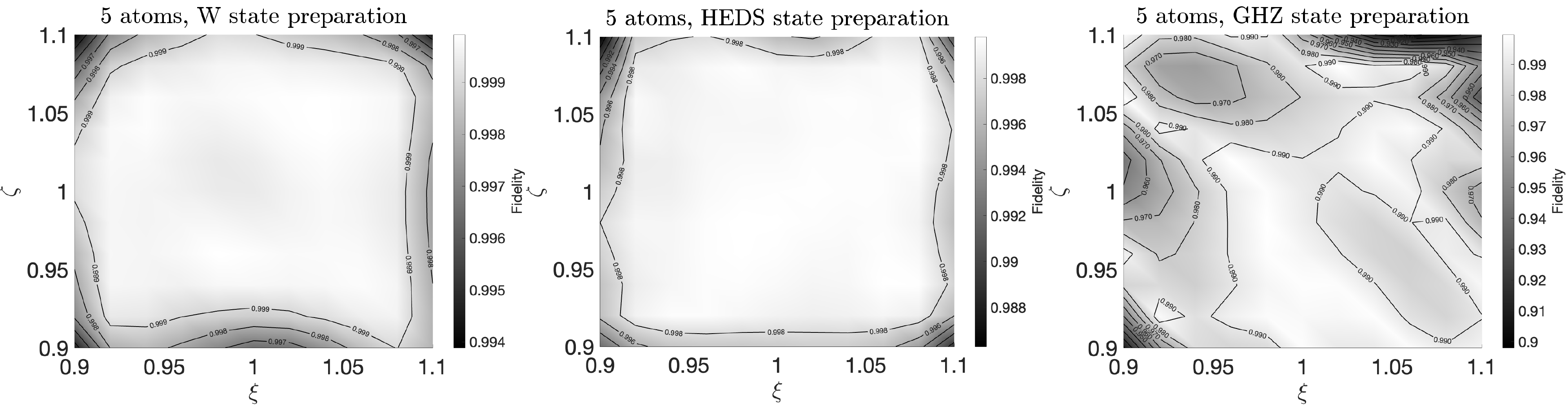}    
\vspace{-4ex}
\caption{Fidelity metric evaluated for designed robust pulses for a population of $N=5$ particles. Designs are for $\xi\in[1-\delta_{\xi},1+\delta_{\xi}]$ and $\zeta\in[1-\delta_{\zeta},1+\delta_{\zeta}]$ with $\delta_{\zeta} = 0.2$ with $\delta_{\zeta} = 0.2$. Desired states to achieve are, from left to right, $|W\rangle$, $|HEDS\rangle$ and $|GHZ\rangle$.}
\label{fig:Results_Fidelity_2}
\end{center}
\vspace{-4ex}
\end{figure*}

\begin{figure*}
\begin{center}
\includegraphics[width=\textwidth]{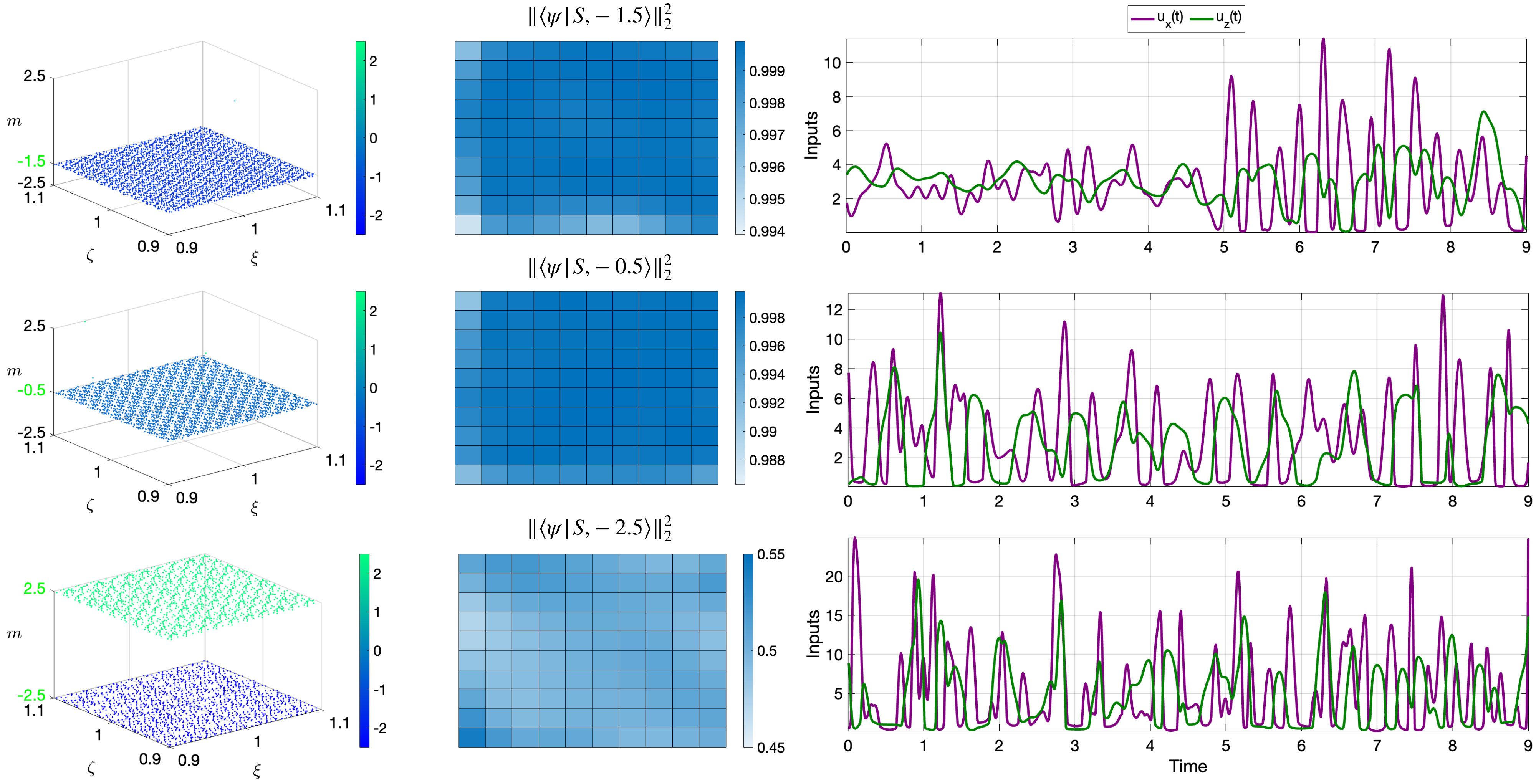}    
\vspace{-4ex}
\caption{Simulated results for the robust design of quantum states for a network with 5 atoms for a system with  $\delta_{\xi} = 0.2$ and $\delta_{\zeta} = 0.2$. Results refer to the design of (from top to bottom) $|W\rangle$, $|HEDS\rangle$ and $|GHZ\rangle$ states. 
The left plots refer to the final achieved states by displaying the density of measured particles in quantum states with eigenvalue $m$ (i.e. ion the $|S, m\rangle$ state) as a function of $\xi$ and $\zeta$, while the second plot displays the total counting of particles in the former. The right plot shows the control profiles obtained for $u_{x}(t)$ and $u_{z}(t)$ for the total time $T=9$.}
\label{fig:Results_Simulations_2}
\end{center}
\vspace{-4ex}
\end{figure*}

The fidelity results for all simulations are presented in Figure \ref{fig:Results_Fidelity}. The results demonstrate that the robust control approach exhibits strong resilience to variations in the inhomogeneity parameters $\xi$ and $\zeta$. Among the target states, the $|W\rangle$ state was the easiest to generate robust pulses for, given its relatively simple structure and proximity to the ground state. Consequently, it achieved fidelity levels exceeding 99\% across all simulations. The $|HEDS\rangle$ state exhibited similarly high performance, with fidelity reductions occurring only in cases with $N=10$ particles and for inhomogeneities approaching $\pm20$\%. However, the $|GHZ\rangle$ state demonstrated a more pronounced fidelity drop below 99\%, reflecting the complexity of its preparation and its greater distance from the ground state. Nonetheless, fidelity stays above 95\% in most cases, aligning with similar results reported in GHZ state preparation studies \cite{Pon:24, Tsu:18}, which however did not compensate for parameter variation in the Hamiltonian.

To further illustrate the obtained results, Figure \ref{fig:Results_Simulations} visualizes the final achieved states for the simulation with $N=10$, $\delta_{\xi}=0.2$ and $\delta_{\zeta}=0$. The results indicate that the final states exhibit a homogeneous profile across all $\xi$ values, with only minor deviations observed in the GHZ state preparation. The control pulses obtained also respect the imposed signal restrictions, demonstrating the feasibility of the optimization framework. Notably, the robust preparation of GHZ states required control pulses with higher amplitudes and sharper peaks compared to the other states, highlighting the increased complexity of achieving high-fidelity GHZ state preparation and the potential for improved results if certain constraints are relaxed.

Overall, the proposed method successfully achieves consistently high fidelity across the ensemble while circumventing the need for extensive sampling of different system configurations, which is a common requirement in alternative robust pulse design approaches. 

\subsection{Robustness in $H_{x}$ and $H_{z}$ Control Hamiltonians} \label{subsec:Results_Double}

The results obtained in Section \ref{subsec:Results_Single} motivate the analysis of the proposed method when considering amplitude discrepancies in both controllers simultaneously, following the complete dynamical evolution dictated by the Hamiltonian in equation \eqref{eq:Hamiltonian_Ensemble}. We preserve all simulation parameters in comparison with the single parameter case, with the exception that now we utilize moments up to the seventh order for both $\xi$ and $\zeta$, which which have uncertainty parametrized by $\delta_{\xi} = \delta_{\zeta}=0.1$. We also alter the initial control profiles to $u_{x,0}(t) = u_{z,0}(t) = 5$. The system is defined by 5 particles, and we attempt to generate an unitary evolution to steer the ensemble from a ground state $|\psi_{0}\rangle$ to one of the desired states described above.

The performance results are displayed through the plots in figure \ref{fig:Results_Fidelity_2}, where high fidelity is achieved for both the c, with the fidelity metric taking values above 99.8\% for most of the parameterized systems and at least 98.6\% in all cases. As in the single control case described in Section \ref{subsec:Results_Single}, preparation of the GHZ state requires the synthesis of controls with the greatest complexity, which is noticeable by how the plots for $|W\rangle$ and $|HEDS\rangle$ states present a homogeneous central area of highest fidelity preparation in comparison with the $|GHZ\rangle$ state.Nonetheless, the fidelity achieved in the GHZ case has reached a lower bound of 96\% for most of the parametrized region, and a overall lower bound of 89.8\%. For a better perspective on the results obtained, one can refer to Table \ref{tab:Double_Inhomogeneity_Indexes}, which displays the maximum, mean and minimum fidelity obtained for each state preparation. We can see all states were able to be prepared with a mean fidelity above 98\%.

The overall performance of the designed pulses is presented in Figure \ref{fig:Results_Simulations_2}. The results indicate that the desired quantum state profiles are consistently achieved across the parameterized interval. This is further supported by the norm profiles of the obtained states, which closely match those of the target preparation states.

However, the pulses depicted in Figure \ref{fig:Results_Simulations_2} highlight notable differences in complexity and amplitude. The preparation of the $|GHZ\rangle$ state required pulses with higher amplitudes and steeper peaks compared to the other target states. To quantitatively assess these differences, we introduce a metric based on the time-integrated absolute value of the control function, as defined in Equation \ref{eq:Index}.

\begin{equation}\label{eq:Index}
    \mathcal{I}_{f(t)} = \int_{0}^{T}|f(t)|dt
\end{equation}

Table \ref{tab:Double_Inhomogeneity_Indexes} presents the computed index values for the control inputs($u_{x}$ and $u_{z}$) and their variation ($\Delta u_{x}$, $\Delta u_{z}$). The results highlight a distinct difference in control complexity, particularly in the variation of control inputs. The increase in these indices reflects the greater challenge associated with preparation of $|GHZ\rangle$ states, emphasizing the higher demands imposed by the more intricate quantum state transitions.

\begin{table}
\centering
\caption{\label{tab:Double_Inhomogeneity_Indexes}Performance data for Section \ref{subsec:Results_Double}}
\begin{tabular}{c c c c}
\hline
\hline
Target State & $|W\rangle$ & $|HEDS\rangle$ & $|GHZ\rangle$\\
\hline
$\max \mathcal{F}(\xi, \zeta, |\psi_{f}\rangle)$ & 0.9999 & 0.9998 & 0.9997\\
$\text{Mean } \mathcal{F}(\xi, \zeta, |\psi_{f}\rangle)$ & 0.9990 & 0.9983 & 0.9808\\
$\min \mathcal{F}(\xi, \zeta, |\psi_{f}\rangle)$ & 0.9939 & 0.9867 & 0.8980\\
$\mathcal{I}_{u_x}$ & 26.6134 & 28.6216 & 47.2684 \\
$\mathcal{I}_{u_z}$ & 27.3416 & 26.2722 & 43.6936 \\
$\mathcal{I}_{\Delta u_x}$ & 2.4042 & 3.6222 & 8.3484\\
$\mathcal{I}_{\Delta u_z}$ & 0.7720 & 1.5900 & 4.1795\\
\hline\hline
\end{tabular}
\vspace{-1ex}
\end{table}

Nonetheless, the performance achieved highlights the effective preparation of quantum states under light pulse amplitude uncertainty, indicating a highly robust design capabilities. Furthermore, the results obtained in this paper can be finely tuned by adjusting parameters such as algorithm related convergence parameters (maximum number of iterations, error target, etc...) as well as the order of moment used, which would improve the representation accuracy of the original parameterized problem.

\section{Conclusion}\label{sec:Conclusion}

We presented a methodology for the robust state preparation of quantum states in a symmetric spin network governed by Ising dynamics and evolving within the Dicke basis. By redefining the problem through a parameterized set of dynamics, we accounted for uncertainties in photonic pulse amplitudes. This reformulation enabled the decomposition of the system’s evolution using Legendre polynomials and facilitated a dual interpretation via moment-based representations, significantly reducing the computational complexity of the large-scale optimization problem.

Integrating this framework into an iterative optimization algorithm for bilinear control systems, we successfully designed control pulses that achieved high-fidelity preparation of quantum states of interest for quantum metrology, specifically the W, HEDS and GHZ states. The pulses presented great robustness towards variation in  multiple controlled electromagnetic amplitude, often applied in similar applications. Moreover, the methodology is versatile and can be extended to generate other quantum superpositions beyond those explicitly considered in this study.

\begin{acknowledgments}
This project was supported by the LDRD program and the Center for Nonlinear Studies at Los Alamos National Laboratory.  Research conducted at Los Alamos National Laboratory is done under the auspices of the National Nuclear Security Administration of the U.S. Department of Energy under Contract No. 89233218CNA000001. Report No. LA-UR-25-21838. This project was supported by the Air Force Office of Scientific Research under the award FA9550-21-1-0335. A. L. P. de Lima gratefully acknowledges the funding of the McDonnell International Scholars Academy.
\end{acknowledgments}

\bibliography{Paper_ref}

\begin{thebibliography}{39}%
\makeatletter
\providecommand \@ifxundefined [1]{%
 \@ifx{#1\undefined}
}%
\providecommand \@ifnum [1]{%
 \ifnum #1\expandafter \@firstoftwo
 \else \expandafter \@secondoftwo
 \fi
}%
\providecommand \@ifx [1]{%
 \ifx #1\expandafter \@firstoftwo
 \else \expandafter \@secondoftwo
 \fi
}%
\providecommand \natexlab [1]{#1}%
\providecommand \enquote  [1]{``#1''}%
\providecommand \bibnamefont  [1]{#1}%
\providecommand \bibfnamefont [1]{#1}%
\providecommand \citenamefont [1]{#1}%
\providecommand \href@noop [0]{\@secondoftwo}%
\providecommand \href [0]{\begingroup \@sanitize@url \@href}%
\providecommand \@href[1]{\@@startlink{#1}\@@href}%
\providecommand \@@href[1]{\endgroup#1\@@endlink}%
\providecommand \@sanitize@url [0]{\catcode `\\12\catcode `\$12\catcode
  `\&12\catcode `\#12\catcode `\^12\catcode `\_12\catcode `\%12\relax}%
\providecommand \@@startlink[1]{}%
\providecommand \@@endlink[0]{}%
\providecommand \url  [0]{\begingroup\@sanitize@url \@url }%
\providecommand \@url [1]{\endgroup\@href {#1}{\urlprefix }}%
\providecommand \urlprefix  [0]{URL }%
\providecommand \Eprint [0]{\href }%
\providecommand \doibase [0]{https://doi.org/}%
\providecommand \selectlanguage [0]{\@gobble}%
\providecommand \bibinfo  [0]{\@secondoftwo}%
\providecommand \bibfield  [0]{\@secondoftwo}%
\providecommand \translation [1]{[#1]}%
\providecommand \BibitemOpen [0]{}%
\providecommand \bibitemStop [0]{}%
\providecommand \bibitemNoStop [0]{.\EOS\space}%
\providecommand \EOS [0]{\spacefactor3000\relax}%
\providecommand \BibitemShut  [1]{\csname bibitem#1\endcsname}%
\let\auto@bib@innerbib\@empty
\bibitem [{\citenamefont {Helstrom}(1969)}]{Hel:69}%
  \BibitemOpen
  \bibfield  {author} {\bibinfo {author} {\bibfnamefont {C.~W.}\ \bibnamefont
  {Helstrom}},\ }\bibfield  {title} {\bibinfo {title} {Quantum detection and
  estimation theory},\ }\href@noop {} {\bibfield  {journal} {\bibinfo
  {journal} {Journal of Statistical Physics}\ }\textbf {\bibinfo {volume}
  {1}},\ \bibinfo {pages} {231} (\bibinfo {year} {1969})}\BibitemShut {NoStop}%
\bibitem [{\citenamefont {Guo}\ \emph {et~al.}(2022)\citenamefont {Guo},
  \citenamefont {Liu}, \citenamefont {Li}, \citenamefont {Li}, \citenamefont
  {Gao}, \citenamefont {Qin},\ and\ \citenamefont {Wen}}]{Guo:22}%
  \BibitemOpen
  \bibfield  {author} {\bibinfo {author} {\bibfnamefont {M.}~\bibnamefont
  {Guo}}, \bibinfo {author} {\bibfnamefont {H.}~\bibnamefont {Liu}}, \bibinfo
  {author} {\bibfnamefont {Y.}~\bibnamefont {Li}}, \bibinfo {author}
  {\bibfnamefont {W.}~\bibnamefont {Li}}, \bibinfo {author} {\bibfnamefont
  {F.}~\bibnamefont {Gao}}, \bibinfo {author} {\bibfnamefont {S.}~\bibnamefont
  {Qin}},\ and\ \bibinfo {author} {\bibfnamefont {Q.}~\bibnamefont {Wen}},\
  }\bibfield  {title} {\bibinfo {title} {Quantum algorithms for anomaly
  detection using amplitude estimation},\ }\href@noop {} {\bibfield  {journal}
  {\bibinfo  {journal} {Physica A: Statistical Mechanics and its Applications}\
  }\textbf {\bibinfo {volume} {604}},\ \bibinfo {pages} {127936} (\bibinfo
  {year} {2022})}\BibitemShut {NoStop}%
\bibitem [{\citenamefont {Giovannetti}\ \emph {et~al.}(2011)\citenamefont
  {Giovannetti}, \citenamefont {Lloyd},\ and\ \citenamefont
  {Maccone}}]{Gio:11}%
  \BibitemOpen
  \bibfield  {author} {\bibinfo {author} {\bibfnamefont {V.}~\bibnamefont
  {Giovannetti}}, \bibinfo {author} {\bibfnamefont {S.}~\bibnamefont {Lloyd}},\
  and\ \bibinfo {author} {\bibfnamefont {L.}~\bibnamefont {Maccone}},\
  }\bibfield  {title} {\bibinfo {title} {Advances in quantum metrology},\
  }\href@noop {} {\bibfield  {journal} {\bibinfo  {journal} {Nature photonics}\
  }\textbf {\bibinfo {volume} {5}},\ \bibinfo {pages} {222} (\bibinfo {year}
  {2011})}\BibitemShut {NoStop}%
\bibitem [{\citenamefont {T{\'o}th}\ and\ \citenamefont
  {Apellaniz}(2014)}]{Tot:14}%
  \BibitemOpen
  \bibfield  {author} {\bibinfo {author} {\bibfnamefont {G.}~\bibnamefont
  {T{\'o}th}}\ and\ \bibinfo {author} {\bibfnamefont {I.}~\bibnamefont
  {Apellaniz}},\ }\bibfield  {title} {\bibinfo {title} {Quantum metrology from
  a quantum information science perspective},\ }\href@noop {} {\bibfield
  {journal} {\bibinfo  {journal} {Journal of Physics A: Mathematical and
  Theoretical}\ }\textbf {\bibinfo {volume} {47}},\ \bibinfo {pages} {424006}
  (\bibinfo {year} {2014})}\BibitemShut {NoStop}%
\bibitem [{\citenamefont {Sengijpta}(1995)}]{Seng:95}%
  \BibitemOpen
  \bibfield  {author} {\bibinfo {author} {\bibfnamefont {S.~K.}\ \bibnamefont
  {Sengijpta}},\ }\href@noop {} {\bibinfo {title} {Fundamentals of statistical
  signal processing: Estimation theory}} (\bibinfo {year} {1995})\BibitemShut
  {NoStop}%
\bibitem [{\citenamefont {Poor}(2013)}]{Poor:13}%
  \BibitemOpen
  \bibfield  {author} {\bibinfo {author} {\bibfnamefont {H.~V.}\ \bibnamefont
  {Poor}},\ }\href@noop {} {\emph {\bibinfo {title} {An introduction to signal
  detection and estimation}}}\ (\bibinfo  {publisher} {Springer Science \&
  Business Media},\ \bibinfo {year} {2013})\BibitemShut {NoStop}%
\bibitem [{\citenamefont {Huang}\ \emph {et~al.}(2024)\citenamefont {Huang},
  \citenamefont {Zhuang},\ and\ \citenamefont {Lee}}]{Hua:24}%
  \BibitemOpen
  \bibfield  {author} {\bibinfo {author} {\bibfnamefont {J.}~\bibnamefont
  {Huang}}, \bibinfo {author} {\bibfnamefont {M.}~\bibnamefont {Zhuang}},\ and\
  \bibinfo {author} {\bibfnamefont {C.}~\bibnamefont {Lee}},\ }\bibfield
  {title} {\bibinfo {title} {Entanglement-enhanced quantum metrology: from
  standard quantum limit to heisenberg limit},\ }\href@noop {} {\bibfield
  {journal} {\bibinfo  {journal} {arXiv preprint arXiv:2402.03572}\ } (\bibinfo
  {year} {2024})}\BibitemShut {NoStop}%
\bibitem [{\citenamefont {Xiong}\ \emph {et~al.}(2010)\citenamefont {Xiong},
  \citenamefont {Ma}, \citenamefont {Liu},\ and\ \citenamefont
  {Wang}}]{Xio:10}%
  \BibitemOpen
  \bibfield  {author} {\bibinfo {author} {\bibfnamefont {H.-N.}\ \bibnamefont
  {Xiong}}, \bibinfo {author} {\bibfnamefont {J.}~\bibnamefont {Ma}}, \bibinfo
  {author} {\bibfnamefont {W.-F.}\ \bibnamefont {Liu}},\ and\ \bibinfo {author}
  {\bibfnamefont {X.}~\bibnamefont {Wang}},\ }\bibfield  {title} {\bibinfo
  {title} {Quantum fisher information for superpositions of spin states},\
  }\href@noop {} {\bibfield  {journal} {\bibinfo  {journal} {Quantum
  Information \& Computation}\ }\textbf {\bibinfo {volume} {10}},\ \bibinfo
  {pages} {498} (\bibinfo {year} {2010})}\BibitemShut {NoStop}%
\bibitem [{\citenamefont {Rubin}(2000)}]{Rub:00}%
  \BibitemOpen
  \bibfield  {author} {\bibinfo {author} {\bibfnamefont {M.~H.}\ \bibnamefont
  {Rubin}},\ }\bibfield  {title} {\bibinfo {title} {Entanglement and state
  preparation},\ }\href@noop {} {\bibfield  {journal} {\bibinfo  {journal}
  {Physical Review A}\ }\textbf {\bibinfo {volume} {61}},\ \bibinfo {pages}
  {022311} (\bibinfo {year} {2000})}\BibitemShut {NoStop}%
\bibitem [{\citenamefont {Erhard}\ \emph {et~al.}(2020)\citenamefont {Erhard},
  \citenamefont {Krenn},\ and\ \citenamefont {Zeilinger}}]{Erh:20}%
  \BibitemOpen
  \bibfield  {author} {\bibinfo {author} {\bibfnamefont {M.}~\bibnamefont
  {Erhard}}, \bibinfo {author} {\bibfnamefont {M.}~\bibnamefont {Krenn}},\ and\
  \bibinfo {author} {\bibfnamefont {A.}~\bibnamefont {Zeilinger}},\ }\bibfield
  {title} {\bibinfo {title} {Advances in high-dimensional quantum
  entanglement},\ }\href@noop {} {\bibfield  {journal} {\bibinfo  {journal}
  {Nature Reviews Physics}\ }\textbf {\bibinfo {volume} {2}},\ \bibinfo {pages}
  {365} (\bibinfo {year} {2020})}\BibitemShut {NoStop}%
\bibitem [{\citenamefont {Friedberg}\ and\ \citenamefont
  {Manassah}(2007)}]{Fri:07}%
  \BibitemOpen
  \bibfield  {author} {\bibinfo {author} {\bibfnamefont {R.}~\bibnamefont
  {Friedberg}}\ and\ \bibinfo {author} {\bibfnamefont {J.}~\bibnamefont
  {Manassah}},\ }\bibfield  {title} {\bibinfo {title} {Dicke states and bloch
  states},\ }\href@noop {} {\bibfield  {journal} {\bibinfo  {journal} {Laser
  Physics Letters}\ }\textbf {\bibinfo {volume} {4}},\ \bibinfo {pages} {900}
  (\bibinfo {year} {2007})}\BibitemShut {NoStop}%
\bibitem [{\citenamefont {Saleem}\ \emph {et~al.}(2024)\citenamefont {Saleem},
  \citenamefont {Perlin}, \citenamefont {Shaji},\ and\ \citenamefont
  {Gray}}]{Sal:24}%
  \BibitemOpen
  \bibfield  {author} {\bibinfo {author} {\bibfnamefont {Z.~H.}\ \bibnamefont
  {Saleem}}, \bibinfo {author} {\bibfnamefont {M.}~\bibnamefont {Perlin}},
  \bibinfo {author} {\bibfnamefont {A.}~\bibnamefont {Shaji}},\ and\ \bibinfo
  {author} {\bibfnamefont {S.~K.}\ \bibnamefont {Gray}},\ }\bibfield  {title}
  {\bibinfo {title} {Achieving the heisenberg limit with dicke states in noisy
  quantum metrology},\ }\href@noop {} {\bibfield  {journal} {\bibinfo
  {journal} {Physical Review A}\ }\textbf {\bibinfo {volume} {109}},\ \bibinfo
  {pages} {052615} (\bibinfo {year} {2024})}\BibitemShut {NoStop}%
\bibitem [{\citenamefont {Gross}(2012)}]{Gro:12}%
  \BibitemOpen
  \bibfield  {author} {\bibinfo {author} {\bibfnamefont {C.}~\bibnamefont
  {Gross}},\ }\bibfield  {title} {\bibinfo {title} {Spin squeezing,
  entanglement and quantum metrology with bose--einstein condensates},\
  }\href@noop {} {\bibfield  {journal} {\bibinfo  {journal} {Journal of Physics
  B: Atomic, Molecular and Optical Physics}\ }\textbf {\bibinfo {volume}
  {45}},\ \bibinfo {pages} {103001} (\bibinfo {year} {2012})}\BibitemShut
  {NoStop}%
\bibitem [{\citenamefont {Albertini}\ and\ \citenamefont
  {D’Alessandro}(2018)}]{Alb:18}%
  \BibitemOpen
  \bibfield  {author} {\bibinfo {author} {\bibfnamefont {F.}~\bibnamefont
  {Albertini}}\ and\ \bibinfo {author} {\bibfnamefont {D.}~\bibnamefont
  {D’Alessandro}},\ }\bibfield  {title} {\bibinfo {title} {Controllability of
  symmetric spin networks},\ }\href@noop {} {\bibfield  {journal} {\bibinfo
  {journal} {Journal of Mathematical Physics}\ }\textbf {\bibinfo {volume}
  {59}} (\bibinfo {year} {2018})}\BibitemShut {NoStop}%
\bibitem [{\citenamefont {Chen}\ \emph {et~al.}(2020)\citenamefont {Chen},
  \citenamefont {Zhou}, \citenamefont {Bian}, \citenamefont {Li},\ and\
  \citenamefont {Peng}}]{Chen:20}%
  \BibitemOpen
  \bibfield  {author} {\bibinfo {author} {\bibfnamefont {J.}~\bibnamefont
  {Chen}}, \bibinfo {author} {\bibfnamefont {Y.}~\bibnamefont {Zhou}}, \bibinfo
  {author} {\bibfnamefont {J.}~\bibnamefont {Bian}}, \bibinfo {author}
  {\bibfnamefont {J.}~\bibnamefont {Li}},\ and\ \bibinfo {author}
  {\bibfnamefont {X.}~\bibnamefont {Peng}},\ }\bibfield  {title} {\bibinfo
  {title} {Subspace controllability of symmetric spin networks},\ }\href@noop
  {} {\bibfield  {journal} {\bibinfo  {journal} {Physical Review A}\ }\textbf
  {\bibinfo {volume} {102}},\ \bibinfo {pages} {032602} (\bibinfo {year}
  {2020})}\BibitemShut {NoStop}%
\bibitem [{\citenamefont {Ouyang}\ \emph {et~al.}(2021)\citenamefont {Ouyang},
  \citenamefont {Shettell},\ and\ \citenamefont {Markham}}]{Ouy:21}%
  \BibitemOpen
  \bibfield  {author} {\bibinfo {author} {\bibfnamefont {Y.}~\bibnamefont
  {Ouyang}}, \bibinfo {author} {\bibfnamefont {N.}~\bibnamefont {Shettell}},\
  and\ \bibinfo {author} {\bibfnamefont {D.}~\bibnamefont {Markham}},\
  }\bibfield  {title} {\bibinfo {title} {Robust quantum metrology with explicit
  symmetric states},\ }\href@noop {} {\bibfield  {journal} {\bibinfo  {journal}
  {IEEE Transactions on Information Theory}\ }\textbf {\bibinfo {volume}
  {68}},\ \bibinfo {pages} {1809} (\bibinfo {year} {2021})}\BibitemShut
  {NoStop}%
\bibitem [{\citenamefont {Carrasco}\ \emph {et~al.}(2024)\citenamefont
  {Carrasco}, \citenamefont {Goerz}, \citenamefont {Malinovskaya},
  \citenamefont {Vuleti{\'c}}, \citenamefont {Schleich},\ and\ \citenamefont
  {Malinovsky}}]{Car:24}%
  \BibitemOpen
  \bibfield  {author} {\bibinfo {author} {\bibfnamefont {S.~C.}\ \bibnamefont
  {Carrasco}}, \bibinfo {author} {\bibfnamefont {M.~H.}\ \bibnamefont {Goerz}},
  \bibinfo {author} {\bibfnamefont {S.~A.}\ \bibnamefont {Malinovskaya}},
  \bibinfo {author} {\bibfnamefont {V.}~\bibnamefont {Vuleti{\'c}}}, \bibinfo
  {author} {\bibfnamefont {W.~P.}\ \bibnamefont {Schleich}},\ and\ \bibinfo
  {author} {\bibfnamefont {V.~S.}\ \bibnamefont {Malinovsky}},\ }\bibfield
  {title} {\bibinfo {title} {Dicke state generation and extreme spin squeezing
  via rapid adiabatic passage},\ }\href@noop {} {\bibfield  {journal} {\bibinfo
   {journal} {Physical Review Letters}\ }\textbf {\bibinfo {volume} {132}},\
  \bibinfo {pages} {153603} (\bibinfo {year} {2024})}\BibitemShut {NoStop}%
\bibitem [{\citenamefont {Stojanovi{\'c}}\ and\ \citenamefont
  {Nauth}(2023)}]{Sto:23}%
  \BibitemOpen
  \bibfield  {author} {\bibinfo {author} {\bibfnamefont {V.~M.}\ \bibnamefont
  {Stojanovi{\'c}}}\ and\ \bibinfo {author} {\bibfnamefont {J.~K.}\
  \bibnamefont {Nauth}},\ }\bibfield  {title} {\bibinfo {title} {Dicke-state
  preparation through global transverse control of ising-coupled qubits},\
  }\href@noop {} {\bibfield  {journal} {\bibinfo  {journal} {Physical Review
  A}\ }\textbf {\bibinfo {volume} {108}},\ \bibinfo {pages} {012608} (\bibinfo
  {year} {2023})}\BibitemShut {NoStop}%
\bibitem [{\citenamefont {Ran}\ \emph {et~al.}(2018)\citenamefont {Ran},
  \citenamefont {Shan}, \citenamefont {Shi}, \citenamefont {Yang},
  \citenamefont {Song},\ and\ \citenamefont {Xia}}]{Ran:18}%
  \BibitemOpen
  \bibfield  {author} {\bibinfo {author} {\bibfnamefont {D.}~\bibnamefont
  {Ran}}, \bibinfo {author} {\bibfnamefont {W.-J.}\ \bibnamefont {Shan}},
  \bibinfo {author} {\bibfnamefont {Z.-C.}\ \bibnamefont {Shi}}, \bibinfo
  {author} {\bibfnamefont {Z.-B.}\ \bibnamefont {Yang}}, \bibinfo {author}
  {\bibfnamefont {J.}~\bibnamefont {Song}},\ and\ \bibinfo {author}
  {\bibfnamefont {Y.}~\bibnamefont {Xia}},\ }\bibfield  {title} {\bibinfo
  {title} {High fidelity dicke-state generation with lyapunov control in
  circuit qed system},\ }\href@noop {} {\bibfield  {journal} {\bibinfo
  {journal} {Annals of Physics}\ }\textbf {\bibinfo {volume} {396}},\ \bibinfo
  {pages} {44} (\bibinfo {year} {2018})}\BibitemShut {NoStop}%
\bibitem [{\citenamefont {Saywell}\ \emph {et~al.}(2020)\citenamefont
  {Saywell}, \citenamefont {Carey}, \citenamefont {Belal}, \citenamefont
  {Kuprov},\ and\ \citenamefont {Freegarde}}]{Say:20}%
  \BibitemOpen
  \bibfield  {author} {\bibinfo {author} {\bibfnamefont {J.}~\bibnamefont
  {Saywell}}, \bibinfo {author} {\bibfnamefont {M.}~\bibnamefont {Carey}},
  \bibinfo {author} {\bibfnamefont {M.}~\bibnamefont {Belal}}, \bibinfo
  {author} {\bibfnamefont {I.}~\bibnamefont {Kuprov}},\ and\ \bibinfo {author}
  {\bibfnamefont {T.}~\bibnamefont {Freegarde}},\ }\bibfield  {title} {\bibinfo
  {title} {Optimal control of raman pulse sequences for atom interferometry},\
  }\href@noop {} {\bibfield  {journal} {\bibinfo  {journal} {Journal of Physics
  B: Atomic, Molecular and Optical Physics}\ }\textbf {\bibinfo {volume}
  {53}},\ \bibinfo {pages} {085006} (\bibinfo {year} {2020})}\BibitemShut
  {NoStop}%
\bibitem [{\citenamefont {Saywell}\ \emph {et~al.}(2023)\citenamefont
  {Saywell}, \citenamefont {Carey}, \citenamefont {Light}, \citenamefont
  {Szigeti}, \citenamefont {Milne}, \citenamefont {Gill}, \citenamefont {Goh},
  \citenamefont {Perunicic}, \citenamefont {Wilson}, \citenamefont {Macrae}
  \emph {et~al.}}]{Say:23}%
  \BibitemOpen
  \bibfield  {author} {\bibinfo {author} {\bibfnamefont {J.~C.}\ \bibnamefont
  {Saywell}}, \bibinfo {author} {\bibfnamefont {M.~S.}\ \bibnamefont {Carey}},
  \bibinfo {author} {\bibfnamefont {P.~S.}\ \bibnamefont {Light}}, \bibinfo
  {author} {\bibfnamefont {S.~S.}\ \bibnamefont {Szigeti}}, \bibinfo {author}
  {\bibfnamefont {A.~R.}\ \bibnamefont {Milne}}, \bibinfo {author}
  {\bibfnamefont {K.~S.}\ \bibnamefont {Gill}}, \bibinfo {author}
  {\bibfnamefont {M.~L.}\ \bibnamefont {Goh}}, \bibinfo {author} {\bibfnamefont
  {V.~S.}\ \bibnamefont {Perunicic}}, \bibinfo {author} {\bibfnamefont {N.~M.}\
  \bibnamefont {Wilson}}, \bibinfo {author} {\bibfnamefont {C.~D.}\
  \bibnamefont {Macrae}}, \emph {et~al.},\ }\bibfield  {title} {\bibinfo
  {title} {Enhancing the sensitivity of atom-interferometric inertial sensors
  using robust control},\ }\href@noop {} {\bibfield  {journal} {\bibinfo
  {journal} {Nature Communications}\ }\textbf {\bibinfo {volume} {14}},\
  \bibinfo {pages} {7626} (\bibinfo {year} {2023})}\BibitemShut {NoStop}%
\bibitem [{\citenamefont {Petruhanov}\ and\ \citenamefont
  {Pechen}(2023)}]{Pet:23}%
  \BibitemOpen
  \bibfield  {author} {\bibinfo {author} {\bibfnamefont {V.~N.}\ \bibnamefont
  {Petruhanov}}\ and\ \bibinfo {author} {\bibfnamefont {A.~N.}\ \bibnamefont
  {Pechen}},\ }\bibfield  {title} {\bibinfo {title} {Grape optimization for
  open quantum systems with time-dependent decoherence rates driven by coherent
  and incoherent controls},\ }\href@noop {} {\bibfield  {journal} {\bibinfo
  {journal} {Journal of Physics A: Mathematical and Theoretical}\ }\textbf
  {\bibinfo {volume} {56}},\ \bibinfo {pages} {305303} (\bibinfo {year}
  {2023})}\BibitemShut {NoStop}%
\bibitem [{\citenamefont {Zeng}\ and\ \citenamefont
  {Allgoewer}(2016)}]{Zeng:16}%
  \BibitemOpen
  \bibfield  {author} {\bibinfo {author} {\bibfnamefont {S.}~\bibnamefont
  {Zeng}}\ and\ \bibinfo {author} {\bibfnamefont {F.}~\bibnamefont
  {Allgoewer}},\ }\bibfield  {title} {\bibinfo {title} {A moment-based approach
  to ensemble controllability of linear systems},\ }\href@noop {} {\bibfield
  {journal} {\bibinfo  {journal} {Systems \& Control Letters}\ }\textbf
  {\bibinfo {volume} {98}},\ \bibinfo {pages} {49} (\bibinfo {year}
  {2016})}\BibitemShut {NoStop}%
\bibitem [{\citenamefont {Narayanan}\ \emph {et~al.}(2020)\citenamefont
  {Narayanan}, \citenamefont {Zhang},\ and\ \citenamefont {Li}}]{Nar:20}%
  \BibitemOpen
  \bibfield  {author} {\bibinfo {author} {\bibfnamefont {V.}~\bibnamefont
  {Narayanan}}, \bibinfo {author} {\bibfnamefont {W.}~\bibnamefont {Zhang}},\
  and\ \bibinfo {author} {\bibfnamefont {J.-S.}\ \bibnamefont {Li}},\
  }\bibfield  {title} {\bibinfo {title} {Moment-based ensemble control},\
  }\href@noop {} {\bibfield  {journal} {\bibinfo  {journal} {arXiv preprint
  arXiv:2009.02646}\ } (\bibinfo {year} {2020})}\BibitemShut {NoStop}%
\bibitem [{\citenamefont {De~Lima}\ \emph {et~al.}(2024)\citenamefont
  {De~Lima}, \citenamefont {Harter}, \citenamefont {Martin},\ and\
  \citenamefont {Zlotnik}}]{De:24}%
  \BibitemOpen
  \bibfield  {author} {\bibinfo {author} {\bibfnamefont {A.~L.~P.}\
  \bibnamefont {De~Lima}}, \bibinfo {author} {\bibfnamefont {A.~K.}\
  \bibnamefont {Harter}}, \bibinfo {author} {\bibfnamefont {M.~J.}\
  \bibnamefont {Martin}},\ and\ \bibinfo {author} {\bibfnamefont
  {A.}~\bibnamefont {Zlotnik}},\ }\bibfield  {title} {\bibinfo {title} {Optimal
  ensemble control of matter-wave splitting in bose-einstein condensates},\
  }in\ \href@noop {} {\emph {\bibinfo {booktitle} {2024 American Control
  Conference (ACC)}}}\ (\bibinfo {organization} {IEEE},\ \bibinfo {year}
  {2024})\ pp.\ \bibinfo {pages} {4196--4203}\BibitemShut {NoStop}%
\bibitem [{\citenamefont {Ning}\ \emph {et~al.}(2022)\citenamefont {Ning},
  \citenamefont {De~Lima},\ and\ \citenamefont {Li}}]{Ning:22}%
  \BibitemOpen
  \bibfield  {author} {\bibinfo {author} {\bibfnamefont {X.}~\bibnamefont
  {Ning}}, \bibinfo {author} {\bibfnamefont {A.~L.~P.}\ \bibnamefont
  {De~Lima}},\ and\ \bibinfo {author} {\bibfnamefont {J.-S.}\ \bibnamefont
  {Li}},\ }\bibfield  {title} {\bibinfo {title} {Nmr pulse design using moment
  dynamical systems},\ }in\ \href@noop {} {\emph {\bibinfo {booktitle} {2022
  IEEE 61st Conference on Decision and Control (CDC)}}}\ (\bibinfo
  {organization} {IEEE},\ \bibinfo {year} {2022})\ pp.\ \bibinfo {pages}
  {5167--5172}\BibitemShut {NoStop}%
\bibitem [{\citenamefont {Li}\ \emph {et~al.}(2022)\citenamefont {Li},
  \citenamefont {Zhang},\ and\ \citenamefont {Kuan}}]{Li:22}%
  \BibitemOpen
  \bibfield  {author} {\bibinfo {author} {\bibfnamefont {J.-S.}\ \bibnamefont
  {Li}}, \bibinfo {author} {\bibfnamefont {W.}~\bibnamefont {Zhang}},\ and\
  \bibinfo {author} {\bibfnamefont {Y.-H.}\ \bibnamefont {Kuan}},\ }\bibfield
  {title} {\bibinfo {title} {Moment quantization of inhomogeneous spin
  ensembles},\ }\href@noop {} {\bibfield  {journal} {\bibinfo  {journal}
  {Annual Reviews in Control}\ }\textbf {\bibinfo {volume} {54}},\ \bibinfo
  {pages} {305} (\bibinfo {year} {2022})}\BibitemShut {NoStop}%
\bibitem [{\citenamefont {Vu}\ \emph {et~al.}(2024)\citenamefont {Vu},
  \citenamefont {Singhal}, \citenamefont {Li},\ and\ \citenamefont
  {Zeng}}]{Vu:24}%
  \BibitemOpen
  \bibfield  {author} {\bibinfo {author} {\bibfnamefont {M.}~\bibnamefont
  {Vu}}, \bibinfo {author} {\bibfnamefont {B.}~\bibnamefont {Singhal}},
  \bibinfo {author} {\bibfnamefont {J.-S.}\ \bibnamefont {Li}},\ and\ \bibinfo
  {author} {\bibfnamefont {S.}~\bibnamefont {Zeng}},\ }\bibfield  {title}
  {\bibinfo {title} {Data-driven moment-based control of linear ensemble
  systems},\ }in\ \href@noop {} {\emph {\bibinfo {booktitle} {2024 American
  Control Conference (ACC)}}}\ (\bibinfo {organization} {IEEE},\ \bibinfo
  {year} {2024})\ pp.\ \bibinfo {pages} {5004--5009}\BibitemShut {NoStop}%
\bibitem [{\citenamefont {de~Lima}\ and\ \citenamefont {Li}(2024)}]{De:24_b}%
  \BibitemOpen
  \bibfield  {author} {\bibinfo {author} {\bibfnamefont {A.~L.~P.}\
  \bibnamefont {de~Lima}}\ and\ \bibinfo {author} {\bibfnamefont {J.-S.}\
  \bibnamefont {Li}},\ }\bibfield  {title} {\bibinfo {title} {A moment-based
  kalman filtering approach for estimation in ensemble systems},\ }\href@noop
  {} {\bibfield  {journal} {\bibinfo  {journal} {Chaos: An Interdisciplinary
  Journal of Nonlinear Science}\ }\textbf {\bibinfo {volume} {34}} (\bibinfo
  {year} {2024})}\BibitemShut {NoStop}%
\bibitem [{\citenamefont {Perlin}\ \emph {et~al.}(2020)\citenamefont {Perlin},
  \citenamefont {Qu},\ and\ \citenamefont {Rey}}]{Per:20}%
  \BibitemOpen
  \bibfield  {author} {\bibinfo {author} {\bibfnamefont {M.~A.}\ \bibnamefont
  {Perlin}}, \bibinfo {author} {\bibfnamefont {C.}~\bibnamefont {Qu}},\ and\
  \bibinfo {author} {\bibfnamefont {A.~M.}\ \bibnamefont {Rey}},\ }\bibfield
  {title} {\bibinfo {title} {Spin squeezing with short-range spin-exchange
  interactions},\ }\href@noop {} {\bibfield  {journal} {\bibinfo  {journal}
  {Physical Review Letters}\ }\textbf {\bibinfo {volume} {125}},\ \bibinfo
  {pages} {223401} (\bibinfo {year} {2020})}\BibitemShut {NoStop}%
\bibitem [{\citenamefont {Novo}\ \emph {et~al.}(2015)\citenamefont {Novo},
  \citenamefont {Chakraborty}, \citenamefont {Mohseni}, \citenamefont {Neven},\
  and\ \citenamefont {Omar}}]{Nov:15}%
  \BibitemOpen
  \bibfield  {author} {\bibinfo {author} {\bibfnamefont {L.}~\bibnamefont
  {Novo}}, \bibinfo {author} {\bibfnamefont {S.}~\bibnamefont {Chakraborty}},
  \bibinfo {author} {\bibfnamefont {M.}~\bibnamefont {Mohseni}}, \bibinfo
  {author} {\bibfnamefont {H.}~\bibnamefont {Neven}},\ and\ \bibinfo {author}
  {\bibfnamefont {Y.}~\bibnamefont {Omar}},\ }\bibfield  {title} {\bibinfo
  {title} {Systematic dimensionality reduction for quantum walks: Optimal
  spatial search and transport on non-regular graphs},\ }\href@noop {}
  {\bibfield  {journal} {\bibinfo  {journal} {Scientific reports}\ }\textbf
  {\bibinfo {volume} {5}},\ \bibinfo {pages} {13304} (\bibinfo {year}
  {2015})}\BibitemShut {NoStop}%
\bibitem [{\citenamefont {D{\"u}r}\ \emph {et~al.}(2000)\citenamefont
  {D{\"u}r}, \citenamefont {Vidal},\ and\ \citenamefont {Cirac}}]{Dur:00}%
  \BibitemOpen
  \bibfield  {author} {\bibinfo {author} {\bibfnamefont {W.}~\bibnamefont
  {D{\"u}r}}, \bibinfo {author} {\bibfnamefont {G.}~\bibnamefont {Vidal}},\
  and\ \bibinfo {author} {\bibfnamefont {J.~I.}\ \bibnamefont {Cirac}},\
  }\bibfield  {title} {\bibinfo {title} {Three qubits can be entangled in two
  inequivalent ways},\ }\href@noop {} {\bibfield  {journal} {\bibinfo
  {journal} {Physical Review A}\ }\textbf {\bibinfo {volume} {62}},\ \bibinfo
  {pages} {062314} (\bibinfo {year} {2000})}\BibitemShut {NoStop}%
\bibitem [{\citenamefont {Gustafson}\ \emph {et~al.}(2019)\citenamefont
  {Gustafson}, \citenamefont {Meurice},\ and\ \citenamefont
  {Unmuth-Yockey}}]{Gus:19}%
  \BibitemOpen
  \bibfield  {author} {\bibinfo {author} {\bibfnamefont {E.}~\bibnamefont
  {Gustafson}}, \bibinfo {author} {\bibfnamefont {Y.}~\bibnamefont {Meurice}},\
  and\ \bibinfo {author} {\bibfnamefont {J.}~\bibnamefont {Unmuth-Yockey}},\
  }\bibfield  {title} {\bibinfo {title} {Quantum simulation of scattering in
  the quantum ising model},\ }\href@noop {} {\bibfield  {journal} {\bibinfo
  {journal} {Physical Review D}\ }\textbf {\bibinfo {volume} {99}},\ \bibinfo
  {pages} {094503} (\bibinfo {year} {2019})}\BibitemShut {NoStop}%
\bibitem [{\citenamefont {Wolf}(2000)}]{Wolf:00}%
  \BibitemOpen
  \bibfield  {author} {\bibinfo {author} {\bibfnamefont {W.~P.}\ \bibnamefont
  {Wolf}},\ }\bibfield  {title} {\bibinfo {title} {The ising model and real
  magnetic materials},\ }\href@noop {} {\bibfield  {journal} {\bibinfo
  {journal} {Brazilian Journal of Physics}\ }\textbf {\bibinfo {volume} {30}},\
  \bibinfo {pages} {794} (\bibinfo {year} {2000})}\BibitemShut {NoStop}%
\bibitem [{\citenamefont {Narayanan}\ \emph {et~al.}(2024)\citenamefont
  {Narayanan}, \citenamefont {Zhang},\ and\ \citenamefont {Li}}]{Nar:24}%
  \BibitemOpen
  \bibfield  {author} {\bibinfo {author} {\bibfnamefont {V.}~\bibnamefont
  {Narayanan}}, \bibinfo {author} {\bibfnamefont {W.}~\bibnamefont {Zhang}},\
  and\ \bibinfo {author} {\bibfnamefont {J.-S.}\ \bibnamefont {Li}},\
  }\bibfield  {title} {\bibinfo {title} {Duality of ensemble systems through
  moment representations},\ }\href@noop {} {\bibfield  {journal} {\bibinfo
  {journal} {IEEE Transactions on Automatic Control}\ } (\bibinfo {year}
  {2024})}\BibitemShut {NoStop}%
\bibitem [{\citenamefont {Folland}(1999)}]{Fol_99}%
  \BibitemOpen
  \bibfield  {author} {\bibinfo {author} {\bibfnamefont {G.~B.}\ \bibnamefont
  {Folland}},\ }\href@noop {} {\emph {\bibinfo {title} {Real analysis: modern
  techniques and their applications}}},\ Vol.~\bibinfo {volume} {40}\ (\bibinfo
   {publisher} {John Wiley \& Sons},\ \bibinfo {year} {1999})\BibitemShut
  {NoStop}%
\bibitem [{\citenamefont {Baker}\ \emph {et~al.}(2024)\citenamefont {Baker},
  \citenamefont {de~Lima}, \citenamefont {Zlotnik},\ and\ \citenamefont
  {Li}}]{Bak:24}%
  \BibitemOpen
  \bibfield  {author} {\bibinfo {author} {\bibfnamefont {L.~S.}\ \bibnamefont
  {Baker}}, \bibinfo {author} {\bibfnamefont {A.~L.~P.}\ \bibnamefont
  {de~Lima}}, \bibinfo {author} {\bibfnamefont {A.}~\bibnamefont {Zlotnik}},\
  and\ \bibinfo {author} {\bibfnamefont {J.-S.}\ \bibnamefont {Li}},\
  }\bibfield  {title} {\bibinfo {title} {{Convergence of Iterative Quadratic
  Programming for Robust Fixed-Endpoint Transfer of Bilinear Systems}},\ }in\
  \href@noop {} {\emph {\bibinfo {booktitle} {63nd IEEE Conference on Decision
  and Control (CDC)}}}\ (\bibinfo {organization} {IEEE},\ \bibinfo {year}
  {2024})\ pp.\ \bibinfo {pages} {8740--8747}\BibitemShut {NoStop}%
\bibitem [{\citenamefont {Pont}\ \emph {et~al.}(2024)\citenamefont {Pont},
  \citenamefont {Corrielli}, \citenamefont {Fyrillas}, \citenamefont {Agresti},
  \citenamefont {Carvacho}, \citenamefont {Maring}, \citenamefont {Emeriau},
  \citenamefont {Ceccarelli}, \citenamefont {Albiero}, \citenamefont
  {Dias~Ferreira} \emph {et~al.}}]{Pon:24}%
  \BibitemOpen
  \bibfield  {author} {\bibinfo {author} {\bibfnamefont {M.}~\bibnamefont
  {Pont}}, \bibinfo {author} {\bibfnamefont {G.}~\bibnamefont {Corrielli}},
  \bibinfo {author} {\bibfnamefont {A.}~\bibnamefont {Fyrillas}}, \bibinfo
  {author} {\bibfnamefont {I.}~\bibnamefont {Agresti}}, \bibinfo {author}
  {\bibfnamefont {G.}~\bibnamefont {Carvacho}}, \bibinfo {author}
  {\bibfnamefont {N.}~\bibnamefont {Maring}}, \bibinfo {author} {\bibfnamefont
  {P.-E.}\ \bibnamefont {Emeriau}}, \bibinfo {author} {\bibfnamefont
  {F.}~\bibnamefont {Ceccarelli}}, \bibinfo {author} {\bibfnamefont
  {R.}~\bibnamefont {Albiero}}, \bibinfo {author} {\bibfnamefont {P.~H.}\
  \bibnamefont {Dias~Ferreira}}, \emph {et~al.},\ }\bibfield  {title} {\bibinfo
  {title} {High-fidelity four-photon ghz states on chip},\ }\href@noop {}
  {\bibfield  {journal} {\bibinfo  {journal} {npj Quantum Information}\
  }\textbf {\bibinfo {volume} {10}},\ \bibinfo {pages} {50} (\bibinfo {year}
  {2024})}\BibitemShut {NoStop}%
\bibitem [{\citenamefont {Tsujimoto}\ \emph {et~al.}(2018)\citenamefont
  {Tsujimoto}, \citenamefont {Tanaka}, \citenamefont {Iwasaki}, \citenamefont
  {Ikuta}, \citenamefont {Miki}, \citenamefont {Yamashita}, \citenamefont
  {Terai}, \citenamefont {Yamamoto}, \citenamefont {Koashi},\ and\
  \citenamefont {Imoto}}]{Tsu:18}%
  \BibitemOpen
  \bibfield  {author} {\bibinfo {author} {\bibfnamefont {Y.}~\bibnamefont
  {Tsujimoto}}, \bibinfo {author} {\bibfnamefont {M.}~\bibnamefont {Tanaka}},
  \bibinfo {author} {\bibfnamefont {N.}~\bibnamefont {Iwasaki}}, \bibinfo
  {author} {\bibfnamefont {R.}~\bibnamefont {Ikuta}}, \bibinfo {author}
  {\bibfnamefont {S.}~\bibnamefont {Miki}}, \bibinfo {author} {\bibfnamefont
  {T.}~\bibnamefont {Yamashita}}, \bibinfo {author} {\bibfnamefont
  {H.}~\bibnamefont {Terai}}, \bibinfo {author} {\bibfnamefont
  {T.}~\bibnamefont {Yamamoto}}, \bibinfo {author} {\bibfnamefont
  {M.}~\bibnamefont {Koashi}},\ and\ \bibinfo {author} {\bibfnamefont
  {N.}~\bibnamefont {Imoto}},\ }\bibfield  {title} {\bibinfo {title}
  {High-fidelity entanglement swapping and generation of three-qubit ghz state
  using asynchronous telecom photon pair sources},\ }\href@noop {} {\bibfield
  {journal} {\bibinfo  {journal} {Scientific reports}\ }\textbf {\bibinfo
  {volume} {8}},\ \bibinfo {pages} {1446} (\bibinfo {year} {2018})}\BibitemShut
  {NoStop}%
\end{thebibliography}%

\end{document}